\documentclass{article}
\usepackage{spconf,amsmath,graphicx,hyperref}
\usepackage{booktabs} 

\title{SPO-CLAPScore: \\Enhancing CLAP-based alignment prediction system with Standardize Preference Optimization, for the first XACLE Challenge}
\name{
Taisei Takano, Ryoya Yoshida
}
\address{The University of Tokyo, Japan\\
\{takano-taisei953, 336921950\}@g.ecc.u-tokyo.ac.jp
}
\begin{document}
%
\maketitle
\begin{abstract}
The first XACLE Challenge (x-to-audio alignment challenge) addresses the critical need for automatic evaluation metrics that correlate with human perception of audio–text semantic alignment.
In this paper, we describe the ``Takano\_UTokyo\_03'' system submitted to XACLE Challenge.
Our approach leverages a CLAPScore-based architecture integrated with a novel training method called Standardized Preference Optimization (SPO).
SPO standardizes the raw alignment scores provided by each listener, enabling the model to learn relative preferences and mitigate the impact of individual scoring biases. Additionally, we employ listener screening to exclude listeners with inconsistent ratings.
Experimental evaluations demonstrate that both SPO and listener screening effectively improve the correlation with human judgment.
Our system achieved 6th place in the challenge with a Spearman's rank correlation coefficient (SRCC) of 0.6142, demonstrating competitive performance within a marginal gap from the top-ranked systems.
The code is available at \url{https://github.com/ttakano398/SPO-CLAPScore}.

\end{abstract}
\section{Introduction}
\label{sec:intro}
Text-to-audio (TTA) generation has become a significant research area due to its ability to synthesize audio samples based on text prompts \cite{majumder2024tango}.
As TTA models generate audio from text input, the key aspect in evaluating these models is to observe the semantic alignment between the text prompt and the generated audio.

Currently, human subjective evaluation remains the gold standard for assessing this aspect, but it is costly in terms of both time and resources.
Although an objective evaluation metric called CLAPScore \cite{huang2023make} is commonly used in the TTA field, it has been reported to exhibit a low correlation with human subjective assessments \cite{Takano_APSIPA2025_01}.
Several studies have addressed the task of creating an automatic evaluation method that correlates with human subjective evaluations of audio–text semantic alignment \cite{Takano_APSIPA2025_01, kanamori25interspeech_relate, kishi26aaai_audiobertscore, deshmukh24b_interspeech, wang2025audioeval}.
However, there have been few established platforms for the unified evaluation of each evaluation method.

XACLE Challenge \cite{XACLE2026} has been launched this year with the aim of developing an automatic evaluation model of audio and text that highly correlates with human subjective evaluations.
This challenge provides XACLE dataset and the baseline model based on LSTM score predictor \cite{kanamori25interspeech_relate}.

In this paper, we present our score prediction system submitted to XACLE Challenge. Our system adopts a CLAPScore-based architecture that leverages the cosine similarity between audio and text embeddings, integrated with a novel optimization method called Standardized Preference Optimization (SPO).
Experimental evaluations on XACLE test dataset demonstrate that SPO successfully enables the model to align its predictions more closely with human judgment.
Furthermore, by ensembling models trained under different conditions, we achieved a robust performance in predicting audio–text alignment scores.

\section{XACLE Challenge}\label{sec:xacle}
XACLE Challenge (x-to-audio alignment challenge) is a competitive challenge that focuses on developing an automatic score evaluation system that assesses the semantic alignment between audio and text \cite{XACLE2026}.
The system's performance is measured using the correlation coefficients and score error between predicted scores and the average semantic-alignments.
The metrics employed include the linear correlation coefficient (LCC), Spearman's rank correlation coefficient (SRCC), and Kendall's rank correlation coefficient (KTAU), and mean squared error (MSE).
Among the four metrics, SRCC was considered the primary performance indicator during the challenge.

The provided dataset contains audio–text pairs, each annotated by four listeners with an 11-point semantic-alignment score, ranging from 0 (lowest) to 10 (highest).
In the training and validation data, three different audio samples were included for each unique text prompt.
It is notable that every listener provided scores for all three audio samples corresponding to the same text.
The listeners for the test data were different from those annotated the training and validation data.
The dataset comprises 7,500 audio–text pairs for the training data and 3,000 pairs for the validation data, 3,000 pairs for the test data.

\section{Proposed Method}\label{sec:proposed}
We propose SPO-CLAPScore, a combination of a CLAPScore-based score prediction model and a preference-based optimization method called Standardized Preference Optimization (SPO).
We will describe the basic architecture, data processing method, and the loss function utilized in our proposed method.

\subsection{CLAPScore based architecture}\label{subsec:architecture}
\begin{figure}[t]
    \centering
    \includegraphics[width=0.45\textwidth]{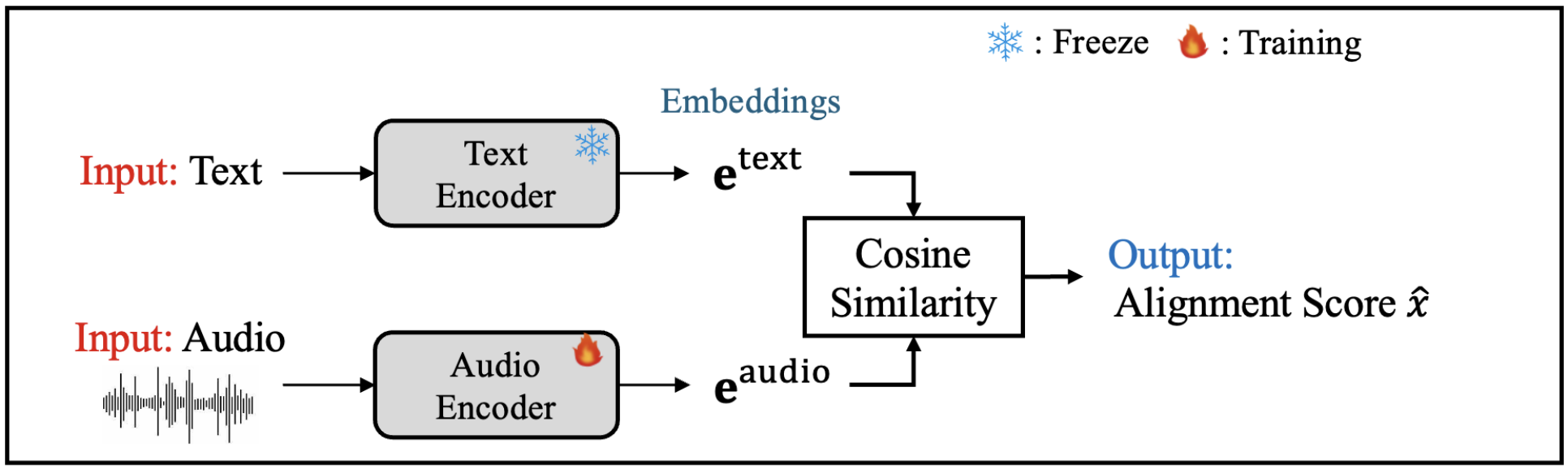}
    \caption{Overview architecture of proposed method}
    \vspace{-8pt}
    \label{fig:proposed}
\end{figure}

The basic architecture of our SPO-CLAPScore follows the framework of Human-CLAP \cite{Takano_APSIPA2025_01} and the CLAPScore \cite{huang2023make}.
Fig.~\ref{fig:proposed} illustrates the overall architecture of the proposed method.

CLAPScore is a metric used in TTA field that measures the alignment between audio and text by calculating the cosine similarity between their CLAP \cite{CLAP2022} embeddings.
Similarly, we calculated the alignment score $\hat{x}$ based on the cosine similarity between the audio and text embeddings:
\begin{equation}
    \hat{x}=\frac{\textbf{e}^{\textsf{audio}}\cdot \textbf{e}^{\textsf{text}}}{\|\textbf{e}^{\textsf{audio}}\| \|\textbf{e}^{\textsf{text}}\|} \times 10, \label{eq:clapscore}
\end{equation}
where $\textbf{e}^{\textsf{text}}$ and $\textbf{e}^{\textsf{audio}}$ denote the output embeddings of the text and audio encoders, respectively.
As the target score is an 11-point score ranging from 0 to 10, we multiplied the score by 10 to adjust the scale.

\subsection{Data processing}\label{subsubsec:process}
To deal with the effect of noisy scores and the individual differences in the data, we performed two types of data processing on XACLE dataset: listener screening and SPO.

\subsubsection{Listener screening}\label{subsubsec:screening}
We found that XACLE dataset contains some scores that are inconsistent with scores annotated by the other listeners.
Because the ground-truth alignment score is calculated by averaging the scores of only four listeners, the effect of these scores can be quite large.
For example, the audio file ``01200.wav'' in XACLE training dataset received a set of scores: $(0, 8, 9, 10)$. 
Although three out of the four listeners scored the data at 8 or higher, the average score was pulled down to $6.75$ due to the single score of $0$.

To address this issue, we filtered the listener based on the algorithm $\Pi(\tau, r)$ described below.
\begin{enumerate}
    \item An individual raw score $x$ is classified as an "NG-Score" if no other score for the same data is included in the interval $[x-\tau, x+\tau]$.
    \item Listeners whose rate of ``NG-Score'' exceeded the threshold $r$ are excluded.
\end{enumerate}

We expect that this listener screening method will exclude noisy scores, enabling the model to learn more fundamental patterns of the alignment scores.

\subsubsection{Standardized Preference Optimization (SPO)}\label{subsubsec:standard}
Dealing with the effect of listener attributes is a key point when we want to predict human evaluated scores automatically.
Instead of incorporating ``listener IDs'' or ``listener contributions'' into the main model, as was done in previous MOS prediction models \cite{baba2024t05, leng2021mbnet}, we propose using a preference based optimization method called Standardized Preference Optimization (SPO) to mitigate the effect of listener attributes when training the model.

\begin{figure}[t]
    \centering
    \includegraphics[width=0.45\textwidth]{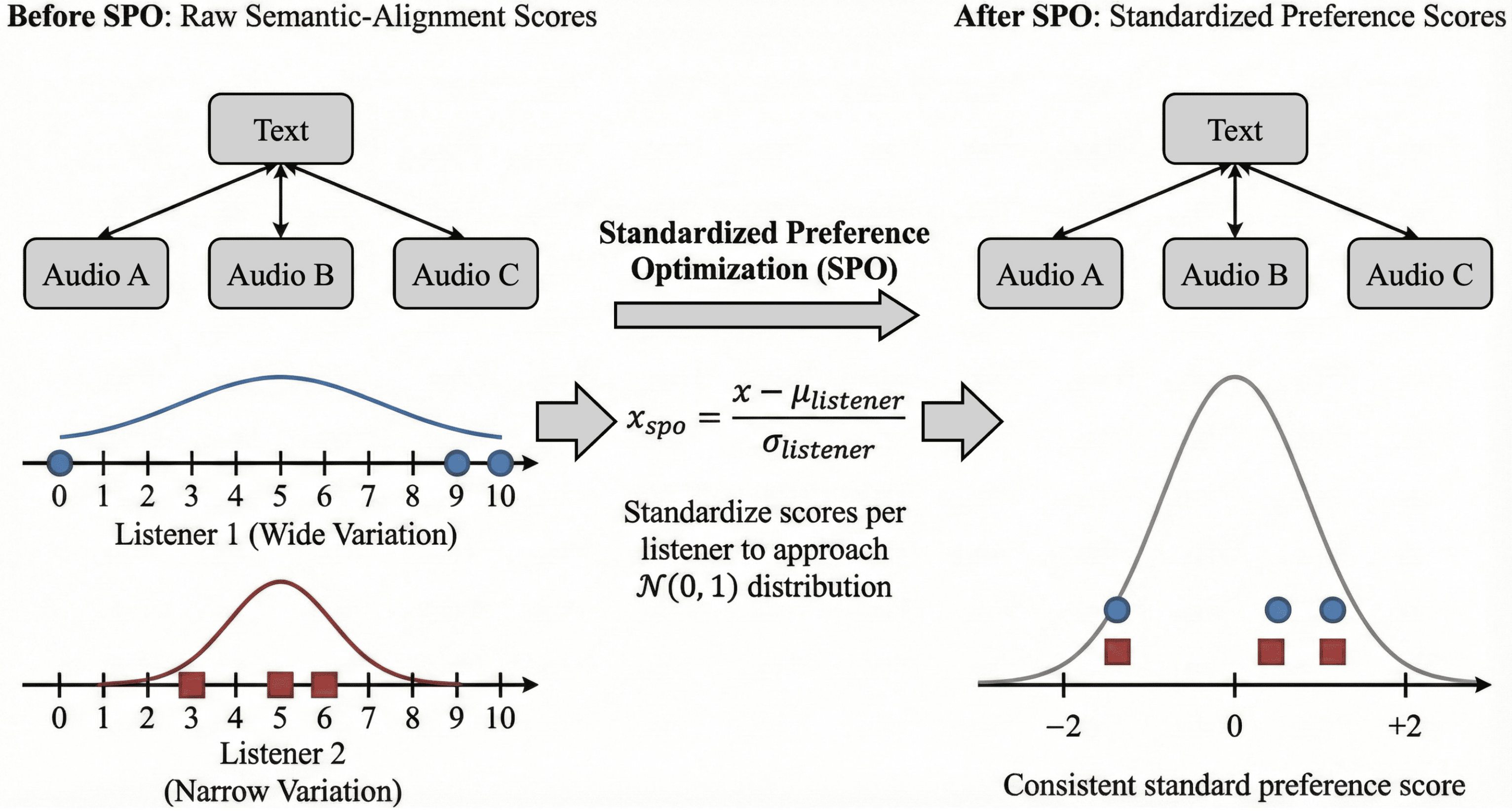}
    \caption{Overview of the Standardized Preference Optimization (SPO)}
    \vspace{-8pt}
    \label{fig:spo}
\end{figure}

The overview of SPO is illustrated on Fig.~\ref{fig:spo}.
As explained in Section \ref{sec:xacle}, the training and validation data of XACLE dataset contains three different audio samples for each text, and every listener provided scores for all three audio samples.
Leveraging this paired structure, we standardized the listener's scores into a ``standard preference score'' to mitigate individual scoring behaviors.

Human annotators often exhibit inherent biases in scoring.
For instance, some listeners may tend to give extreme scores (e.g., frequently assigning 0 or 10), while others give conservative scores, restricting their scores to a narrower central range (e.g., staying between 3 and 8).
Learning directly from these raw semantic-alignments allows such personal variances to confuse the model.
SPO addresses this by transforming the raw semantic-alignments into relative indicators.
A positive $x_{\text{spo}}$ signifies that the sample was rated higher than the listener's personal average, regardless of the raw numerical value.
This transformation enables the model to disregard the noisy scale of raw data and focus on capturing the human preference patterns.

We standardize the raw semantic-alignment scores for each listener using the equation below, ensuring the resulting score distribution follows $\mathcal{N}(0, 1)$:
\begin{equation}
    x_{\text{spo}}=\frac{x-\mu_{\text{listener}}}{\sigma_{\text{listener}}},
\end{equation}
where $\mu_{\text{listener}}$ and $\sigma_{\text{listener}}$ denote the mean and the standard deviation of all scores provided by the same listener across the dataset.
$x_{\text{spo}}$ denotes the ``standard preference score'' and we used this score during the loss calculation to optimize our model, instead of using raw semantic-alignment scores.

\subsection{Loss function}\label{subsec:loss}
Our model is trained to minimize a loss function that combines the regression loss $L_{\text{reg}}$ and the contrastive loss $L_{\text{con}}$ \cite{saeki22c_interspeech}.
We utilized the mean squared error (MSE) between the predicted score and the target score as the regression loss $L_{\text{reg}}$.

Since we employed the standardized preference scores by SPO for optimizing the model, we normalized the predicted scores during loss calculation using the global mean $\mu_{\text{train}}$ and standard deviation $\sigma_{\text{train}}$ of the raw semantic-alignment scores in the training dataset. The overall loss function is defined as below:

\begin{align}
L =& L_{\text{reg}}\left(x_{\text{spo}}, \frac{\hat{x}-\mu_{\text{train}}}{\sigma_{\text{train}}}\right) \nonumber \\
 {} &+ \lambda L_{\text{con}}\left(x_{\text{spo}}, \frac{\hat{x}-\mu_{\text{train}}}{\sigma_{\text{train}}}\right),
\end{align}
where $x_{\text{spo}}$ denotes the target ``standard preference score'', and $\hat{x}$ denotes the predicted score. $\lambda$ is the hyperparameter to weight contrastive loss.

\section{Evaluations}\label{sec:eval}

In this section, we present the detailed experimental results and training configurations of our system, ``Takano\_UTokyo\_03'', submitted to XACLE Challenge.

\subsection{Experimental conditions}\label{subsec:setup}
The submitted system was constructed by ensembling models trained under the specific configurations (Setting A, Setting B, and Setting C) described below:
\begin{itemize}
    \item Setting A: No screening; no contrastive learning loss.
    \item Setting B: With screening; with contrastive learning loss.
    \item Setting C: With screening; no contrastive learning loss.
\end{itemize}
For each of the three settings, we trained models with and without warm-up, using three different random seeds for each case. This resulted in a total of six models per setting.
By combining models trained on both screened and non-screened datasets, we aimed to enhance robustness across varying conditions.
The final prediction score was obtained by averaging the predictions from the individual models.


The common training configurations shared across all models are summarized in Table \ref{tb:set-common}.
By applying listener screening with $\Pi(\tau=5, r=0.2)$ as in Table~\ref{tb:set-common}, the number of raw audio–text alignment scores reduced from 30,000 samples to 29,308 samples in XACLE training dataset, and from 12,000 samples to 11,747 samples in XACLE validation dataset.

We adopted M2D-CLAP 2025 \cite{niizumi2025m2d-clap} as the audio encoder and BERT \cite{DBLP:journals/corr/abs-1810-04805} as the text encoder, after evaluating several backbone encoders \cite{li2024advancing, mei2023wavcaps}. To prevent overfitting, we froze the text encoder and fine-tuned only the audio encoder.
\begin{table}[h]
    \centering
    \caption{Common configurations}
    \label{tb:set-common}
    \begin{tabular}{l c c c}
        \toprule
        Configurations &   \\
        \midrule
        Audio encoder & M2D-CLAP 2025 \\
        Text encoder & BERT (base) \\
        Optimizer & Adam\\
        Total epochs & 50 \\
        Initial learning rate (if warm-up) & 0.0\\
        Peak learning rate (if warm-up) & 0.0001 \\
        Peak epoch (if warm-up) & 5\\
        $\lambda$ (if contrastive) & 0.5\\
        $\tau$ (if screening) & 5\\
        $r$ (if screening) & 0.2\\
        \bottomrule
    \end{tabular}
\end{table}


\subsection{Overall results}
The overall performance results of our SPO-CLAPScore system, submitted to XACLE Challenge, are presented in Table~\ref{tab:main_results}.
These results are sourced from the official XACLE Challenge leaderboard\footnote{https://xacle.org/results.html}.
We confirmed that our models outperform the baseline model across all metrics.
Notably, our main SPO-CLAPScore model achieved the best performance among the submitted variations, improving the SRCC by more than $0.27$ compared to the baseline.
Regarding our ensembling strategy, the results suggest that ensembling models trained under different conditions, such as listener screening and model warmup, enabled the ensemble to capture diverse features and enhance its robustness.

Although the test dataset does not employ our listener screening method, ``SPO-CLAPScore w/o setting A'' (only trained on data with listener screening) outperformed ``SPO-CLAPScore w/o setting B\&C'' (only trained on data without listener screening) in all metrics, especially showing a notable difference in LCC and MSE. Since LCC and MSE are sensitive to outliers, these results suggest that our listener screening method enables the model to successfully generate more stable and robust scores.

\begin{table}[t]
    \centering
    \caption{Evaluation results of the submitted models on XACLE test dataset}\label{tab:main_results}
    \begin{tabular}{l|c c c c}
        \toprule
         & SRCC $\uparrow$ & LCC $\uparrow$ & KTAU $\uparrow$ & MSE $\downarrow$  \\
        \midrule
        Baseline & 0.3345 & 0.3420 & 0.229 & 4.811\\
        \midrule
        SPO-CLAPS & \textbf{0.6142} & \textbf{0.6542} & \textbf{0.4407} & 2.985 \\ 
        w/o Setting B \& C & 0.6118 & 0.6510 & 0.4391 & 3.072 \\ 
        w/o Setting A & 0.6138 & 0.6542 & 0.4401 & \textbf{2.963} \\ 
        \bottomrule
    \end{tabular}
\end{table}

\subsection{Our results on the first XACLE Challenge}
Our system, ``Takano\_UTokyo\_03'', achieved 6th place in the official results of XACLE Challenge.
Notably, the gap between our model and the 5th-place system was marginal, with a difference of only 0.0001 in SRCC.
In contrast, we maintained a significant lead of more than 0.04 points in SRCC over the lower-ranked systems.

\subsection{Ablation study}
We conducted an ablation study to verify the effectiveness of SPO.
In this experiment, we used a single SPO-CLAPScore model trained under Setting B without ensembling, and compared the performance on XACLE validation data with and without SPO.
Since the ground-truth score of XACLE test data is not accessible at present, we used the validation data of XACLE dataset instead for this experiment.

As illustrated in Table~\ref{tab:abu}, we confirmed that the model with SPO outperforms those without SPO across all metrics. This indicates that our SPO effectively enables the model to learn preferences on audio–text semantic alignment closer to human judgment.
\begin{table}[t]
    \centering
    \caption{Ablation on Standardized Preference Optimization on XACLE validation data (no ensembling)}\label{tab:abu}
    \begin{tabular}{l|c c c c} 
        \toprule
         & SRCC $\uparrow$ & LCC $\uparrow$ & KTAU $\uparrow$ & MSE $\downarrow$  \\
        \midrule
        SPO-CLAPS & \textbf{0.6367} & \textbf{0.6572} & \textbf{0.4606} & \textbf{3.256}\\
        w/o SPO & 0.5408 & 0.5514 & 0.3837 & 6.709 \\
        \bottomrule
    \end{tabular}
\end{table}

\section{Conclusion}\label{sec:conclusion}
We presented our ``Takano\_UTokyo\_03'' system submitted to XACLE challenge.
Our system is built upon a CLAPScore-based architecture utilizing the cosine similarity between audio and text embeddings, integrated with a novel optimization method called Standardized Preference Optimization (SPO).
By standardizing the target alignment scores for each listener, SPO enabled the model to learn preferences closer to human judgment while mitigating the effect of individual listener attributes.
Also, by ensembling models trained with different conditions, our model successfully exhibited robust audio–text alignment score, close to human perception.

Experimental evaluations on XACLE datasets demonstrated that our method succeeded to improve the correlation between predicted audio–text semantic-alignment scores and the human evaluations.
Our future work includes constructing more general and robust score prediction system by collecting a wider variety of data.

\bibliographystyle{IEEEbib}
\bibliography{refs}

\end{document}